\documentclass[12pt]{iopart}

\pdfoutput=1

\usepackage{graphicx}
\usepackage{epsfig}
\usepackage{setstack} 
\usepackage{mathenv}
\usepackage{multirow}
\usepackage{amsfonts}
\usepackage{amssymb}
\usepackage{setspace}
\usepackage{color}
\usepackage{subfigure}
\usepackage{notes2bib}
\usepackage[font=small,format=plain,labelfont=bf,up,textfont=normal,up]{caption}

\begin{document}

\title{Evaluation of nitrogen- and silicon-vacancy defect centres as single photon sources in quantum key distribution}

\author{Matthias Leifgen$^{1}$, Tim Schr\"{o}der$^{1,2}$, Friedemann G\"{a}deke$^1$, Robert Riemann$^1$, Valentin M\'{e}tillon$^3$, Elke Neu$^{4,5}$, Christian Hepp$^4$, Carsten Arend$^4$, Christoph Becher$^4$, Kristian Lauritsen$^6$, and Oliver Benson$^1$}
\address{$^1$ Nano-Optik, Institut f\"{u}r Physik, Humboldt-Universit\"{a}t zu Berlin, 12489 Berlin, Germany}
\address{$^2$ present adress: Department of Electrical Engineering and Computer Science, Massachusetts Institute of Technology,
Cambridge, Massachusetts 02139, USA}
\address{$^3$ \'{E}cole normale sup\'{e}rieure, 45 rue d'Ulm, Paris, France}
\address{$^4$ Fachrichtung 7.2 (Experimentalphysik), Universit\"{a}t des Saarlandes, Campus E2.6, 66123 Saarb\"{u}cken, Germany}
\address{$^5$ present adress: Universit\"{a}t Basel, Departement Physik, Klingelberstrasse 82, 4056 Basel, Switzerland}
\address{$^6$ PicoQuant GmbH, Rudower Chaussee 29, 12489 Berlin, Germany}

\ead{\mailto{leifgen@physik.hu-berlin.de,schroder@mit.edu}}


\begin{abstract} 
We demonstrate a quantum key distribution (QKD) testbed for room temperature single photon sources based on defect centres in diamond. A BB84 protocol over a short free-space transmission line is implemented. The performance of nitrogen-vacancy (NV) as well as silicon-vacancy defect (SiV) centres is evaluated and an extrapolation for next-generation sources with enhanced efficiency is discussed.
\end{abstract}

\maketitle %

\tableofcontents

\section{Introduction}
Single photons are a key ingredient in many quantum information processing (QIP) applications. Examples are quantum key distribution (QKD) \cite{Gisin2002}, long distance quantum repeater protocols \cite{Sangouard2007} or linear optical quantum computing. However, in QKD, high key rate and/or long distance experiments have been succesfully implemented without true single photons using weak coherent laser pulses (WCP) \cite{Dixon2010,Schmitt-Manderbach2007} together with the decoy state protocol \cite{Hwang2003,Lo2005}. Although this protocol is secure against photon number splitting (PNS) attacks, it has the disadvantage of producing some overhead, because one explicitly uses vacuum and very low intensity pulses together with the signal pulses and also operates with faint pulses with a mean photon number per pulse of around 0.5. A true single photon source (SPS) with high efficiency could therefore still be favourable compared to WCP with decoy states.
Efficiency means that it emits a single photon into a well defined spectral and spatial mode with a probability near unity each time a trigger is applied. A practical SPS should also have a stable emission rate, be easy-to-use and should operate at room temperature.

The most promising candidates for practical true SPS today are solid-state emitters, such as quantum dots (QDs) \cite{Intallura2007,Heindel2012} or defect centres in diamond \cite{Kurtsiefer2000,Neu2011}. QDs are available for a broad range of wavelengths, including the telecom bands, can be excited optically or electrically and are efficient with coupling rates to a usable output mode of up to over 70\,\% \cite{Claudon2010}. One major drawback of QDs is until now the need for cryogenic cooling.
Regarding defect centres in diamond, the most intensely studied defect centre is the nitrogen-vacancy (NV) centre \cite{Kurtsiefer2000}. It is formed by a substitutional nitrogen atom together with an adjacent vacancy. The NV center occurs in a neutral and a negatively charged state \cite{Siyushev2013}. In the following we only regard the charged NV center. Its electronic structure is well understood \cite{Hossain2008,Manson2006} and can be described by a three level model concerning its optical properties. Optical and electrical (neutral NV) excitation was demonstrated \cite{BeveratosII2002,Mizuochi2012}. Also, another defect centre, the silicon-vacancy centre (SiV) has been studied \cite{Neu2011,Neu2012}. It consists of a silicon impurity in a so called split-vacancy-configuration \cite{Albrecht2013}. One key feature of defect centers is their optical stability and high quantum yield even at room temperature. Whereas emission into the zero phonon line (ZPL) is a few \% for the NV center, the SiV emits 70-80\,\% of photons into the zero phonon line (ZPL). Both for NV and SiV centres high photon generation rates exceeding 2 respectively 6\,Mcps under continous laser excitation \cite{Schroder2011,Neu2011} have been reported. Other, presumably Cr-related centers with potentially even higher count rates have been investigated \cite{Aharonovich2010,AharonovichII2010}, but their reliable fabrication is problematic and a full understanding of their structural properties is missing. So far only the NV centre has been utilized as SPS in a QKD experiment \cite{Beveratos2002,Alleaume2004}.

In order to evaluate the applicability of defect centres as reliable sources for QKD, it is desirable to use test-beds that allow for long-term measurements, implementation of different protocols, as well as straightforward integration of different defect centres. In this paper, we report the realization of such a testbed. It consists of a short free-space transmission line combined with a compact SPS based on defect centres. The source relies on a specialized confocal setup for stable optical excitation and efficient collection of single photons from defect centres. Furthermore, it is designed in a way that it is easy to replace one kind of defect centre by another. QKD experiments are performed with NV centres and for the first time also with SiV centres.

\section{Compact and versatile design of single photon source}
\begin{figure}[h] 
\centering
\includegraphics [width=12cm] {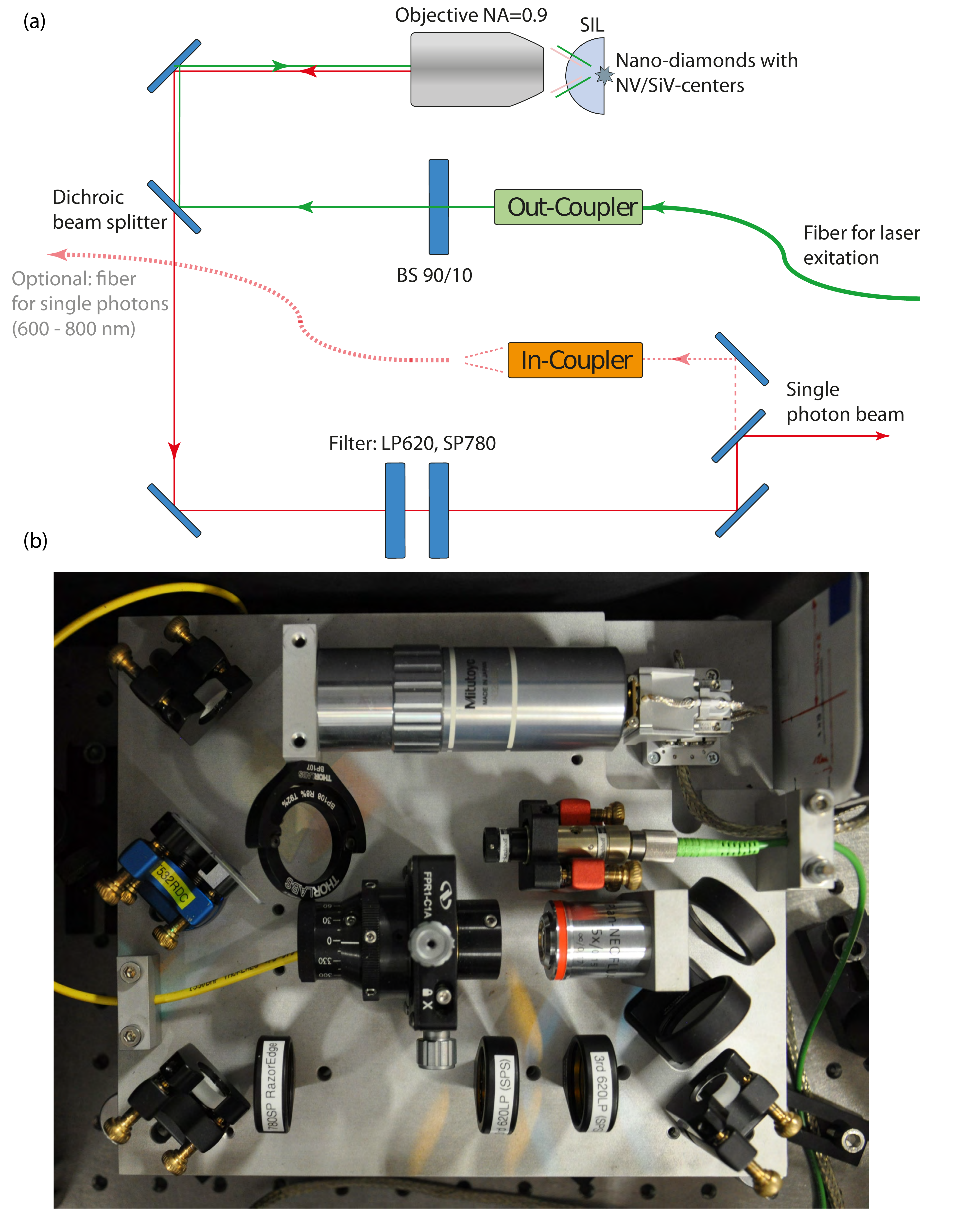}\\
\caption{(a) Scheme and (b) photo of the compact confocal setup. The excitation laser is focused with a high NA objective onto the sample which is either one of several diamonds spincoated on a solid immersion lens (SIL) or grown on a substrat. The emission is collected by the same objective and then filtered by a dichroic beam splitter and longpass (LP) and shortpass (SP) filters to clean it from residual laser light or fluorescence of the SIL or the substrate.}
\label{fig:ConfSetup}
\end{figure}

The design of the source relies on a compact, portable and ready to use confocal setup. A $ZrO_{2}$ solid immersion lens (SIL) can be utilized to enhance the collection of single photons emitted from defect centres in nanodiamonds spin-coated directly on the SILs. Details of the fabrication of SILs with NV centres are provided in \cite{Schroder2011}. The source is built for single photon emitters with a wavelength range from 600 to 800\,nm. With this it allows for the  implementation of SPS using NV, SiV as well as Cr-based defect centres. Figure \ref{fig:ConfSetup} shows a schematic (a) and a photograph (b) of the source which fits completely on an aluminum plate and has dimensions of only $22.5\,cm \times 19\,cm \times 9\,cm$. In this way the SPS is mobile and can easily be integrated in different experimental setups. The setup is robust against mechanical vibrations and thermal drifts due to its small size and compact mounting of all optical components. The generated single photon beam can either be freespace or fiber coupled by removal/addition of a single mirror which is equipped with a magnetic base. The sample unit holding the defect centres can either be the SIL with spin-coated defect centres or another substrate due to a removable sample holder. The setup is equipped with broadband optics and thus suitable for various defect centres, provided their emission wavelength is in the range of 600\,nm to 800\,nm. The sample holder is mounted on a 3-axes piezo stage. In order to keep track of the absolute position of the stage, sensors capable of detecting changes on a nanometer scale are used (SmarAct System). With this system it is possible to focus on a well defined position on the sample with very high accuracy and stability.

\section{Properties of colour centres used in the single photon sources}
In principle, any kind of colour centre can be employed in the source described in the previous section.  In the following, we focus on nitrogen-vacancy (NV) and silicon-vacancy (SiV) centres.

\subsection{Nitrogen-vacancy centres in diamonds as single photon source}
\label{subsec:NV}
NVs have been proven to be stable and bright single photon emitters and have been extensively studied and described elsewhere \cite{Kurtsiefer2000,Rabeau2007,Schroder2011}. They have a broad emission spectrum from 600-800\,nm. High count rates of up to 2.4\,Mcps (cw excitation) and relatively high overall photon yields (the ratio between excitation and detection rates) of up to 4.2\,\% have been achieved \cite{Schroder2011}. The quality of single photon emission, in particular the contribution of multi-photon events is routinely determined by measuring the second order autocorrelation function ($g^{(2)}(\tau)$) in a Hanbury-Brown Twiss (HBT) setup \cite{HanburyBrown1956}. Although in the ideal case $g^{(2)}(\tau=0)$ = 0, a value of $g^{(2)}(\tau=0)$ $\textless$ $0.5$ is generally accepted as a criterion for single photon emission. Values below 0.12 have been reported with NV defect centres \cite{Schroder2011}. In QKD it is favourable to have photons at a well defined instant of time, thus pulsed excitation of the defect centres is of interest. In such a pulsed excitation scheme, using a green diode laser (PicoQuant LDH-P-FA-530 \cite{Schoenau2011}, 531\;nm, pulse width \textless 100\,ps), we achieved count rates of 8900\,cps at an excitation rate of 1\,MHz for an NV centre in a nanodiamond which was spincoated on a SIL. This corresponds to an overall photon yield of 0.89\,\% and a source efficiency of 2.9\,\%. The latter is defined as the ratio of excitation pulses resulting in a single photon without background in the desired optical mode, here the freespace beam of the QKD experiment. This value is determined for a given overall photon yield by taking the overall transmission $\eta_{setup}$ of 0.31 of our setup, including the efficiency of $\sim65\,\%$ of the avalanche photodetectors (APDs), into account. For the QKD experiment, the maximal excitation rate was limited to frequencies up to 1\,MHz by the modulation rate of the electro optic modulators (EOMs) (see section \ref{sec:QKDsetup}). A $g^{(2)}(0)$ value under pulsed excitation of 0.09, clearly indicating high purity single photon emission, is measured (fig. \ref{fig:gtwo}a) using high resolution time-correlation electronics (PicoHarp 300 from PicoQuant). A lifetime of 28.5$\pm$1.5\,ns is calculated  from the pulse shape.

\subsection{Silicon-vacancy centres in diamonds as single photon source} 
\label{subsec:SiV}
\begin{figure}[h] \center
\includegraphics [width=15cm] {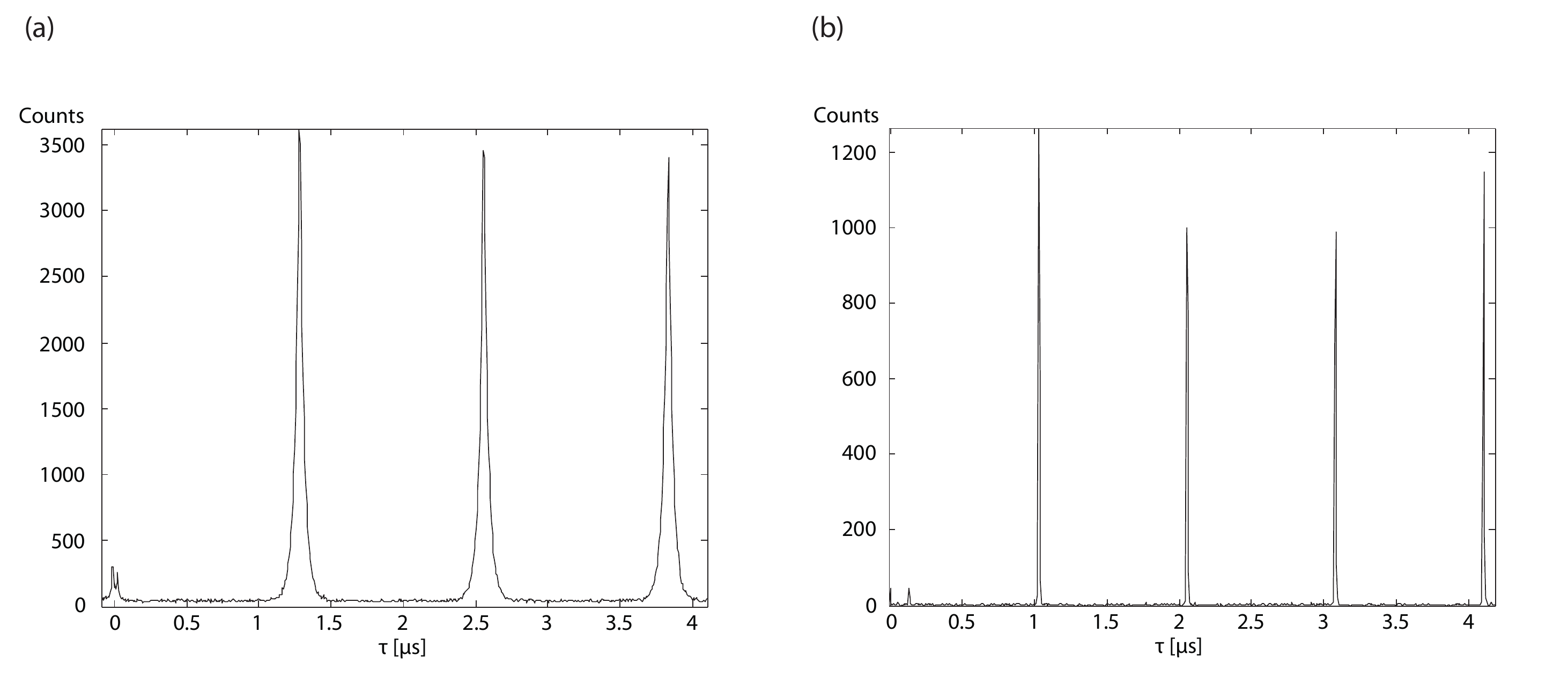}\\
\caption{Measured intensities as a function of time for NV (a) and SiV (b) emission under pulsed excitation to calculate $g^{(2)}(\tau)$. The excitation rates have been 800\,kHz and 1\,MHz, respectively. The missing pulse at $\tau = 0$ indicates single photon emission. From the pulse shape we calculated al lifetime of the excited state of 28.5$\pm$1.5\,ns for the NV and 3$\pm$2\,ns for the SiV.} 
\label{fig:gtwo}
\end{figure}

A disadvantage of the NVs is their phonon-broaded emission spectrum and long lifetime in the order of 10\,ns, which limits the maximal excitation and emission rate. SiVs have a much narrower, linearly polarized emission spectrum of about 1-2\,nm at around 739\,nm \cite{Neu2011}, due to lower phonon coupling and a concentration of the emission into the ZPL, and a shorter lifetime in the order of 1\,ns, making them an interesting candidate for QKD with single photons, especially for freespace daylight operations \cite{Buttler2000}. There have been reports on high photon emission rates of up to 6.2\,Mcps using cw excitation and values of $g^{(2)}(0)$ below 0.05 \cite{Neu2012}. Single SiV centers are created during the chemical vapor deposition (CVD) growth of randomly oriented nanodiamonds (NDs)  on Iridium (Ir) films \cite{Neu2011}.
We excite our SiV sample with laser pulses from a red diode laser (PicoQuant diode laser LDH-D-C-690, 687\,nm, pulse width \textless 100\,ps) at an excitation frequency of 1\,MHz. With the brightest SiV centre we achieved a photon count rate of 3700\,cps and a $g^{(2)}(0)$ value of 0.04, see fig. \ref{fig:gtwo}b. From the pusle shape, a lifetime of 3$\pm$2\,ns is calculated. The overall photon yield was 0.37\,\% and the plain source efficiency is thus 1.2\,\%. The achievable count rate per excitation pulse was lower for the SiV centre compared to the NV centre. Knowing that the collection efficiency of emitted photons from SiV nanodiamonds grown on Ir-substrate can be very high \cite{Neu2012}, this hints a lower quantum efficiency. In \cite{Neu2012}, a quantum efficiency between 1-9\,\% was estimated. Compared to the NV centre, the excitation frequency could be chosen to be much higher for the SiV centre due to the shorter lifetime, which could compensate for the lower quantum efficiency (cf. section \ref{sec:discussion})

It is also observed that SiVs do not have the same stability as NVs under excitation. Some of them bleach, probably due to photo-ionization, especially when excited at higher excitation powers. It has been suggested to circumvent bleaching by creating free electrons close to the sample by shining a blue laser at its vicinity \cite{Arend2011}. Using surface treatment or a better control of the impurity content of the samples might also be a way to produce more photostable SiVs in the future \cite{Neu2012}.

\section{Setup of QKD testbed}
\label{sec:QKDsetup}
\begin{figure}[h] \center
\includegraphics [width=15cm] {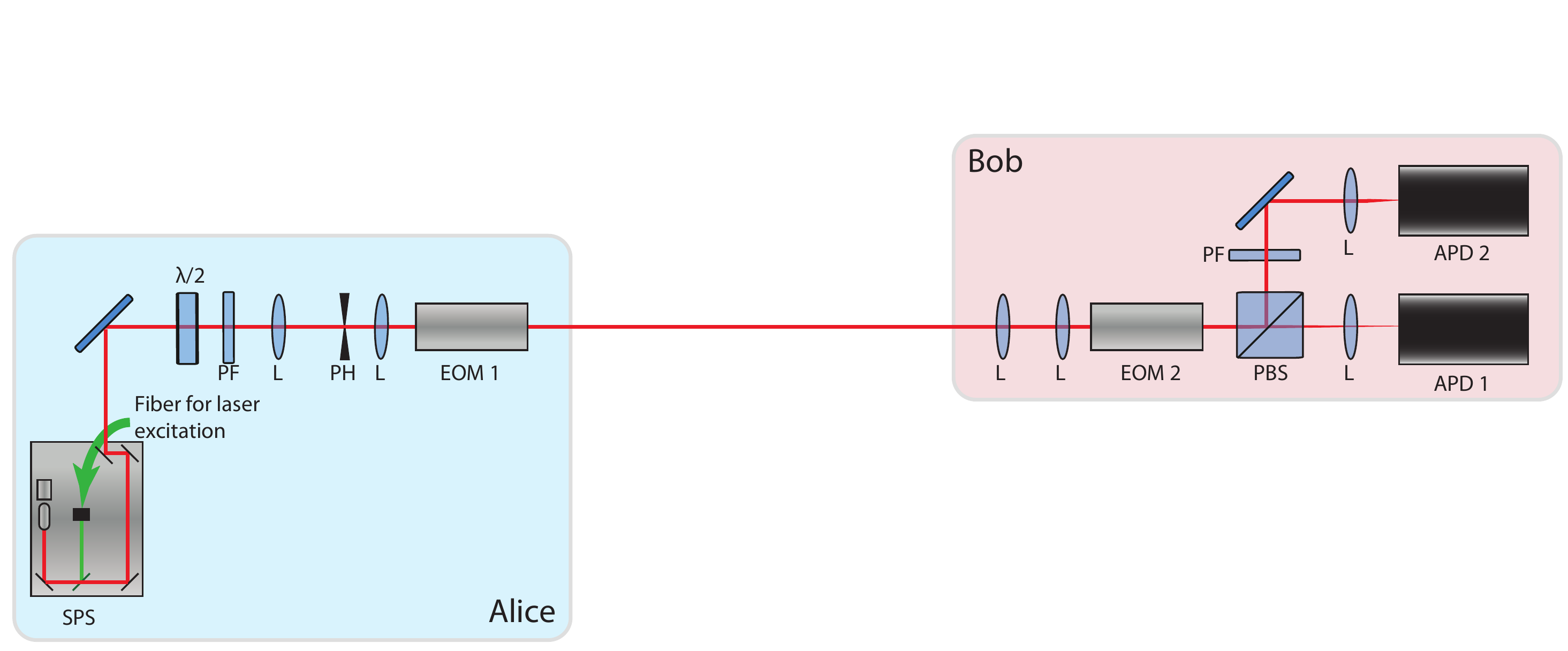}\\
\caption{Schematics of the QKD testbed. Single photons are emitted from the source (SPS) and prepared in a well defined polarization state by the $\lambda/2$ plate and a polarization filter (PF). After collimation and spatial mode cleaning by two lenses (L) and a pinhole (PH), they pass through an electro-optical modulator (EOM 1). On Bob's side, the beam is recollimated by two lenses and then passes through a second modulator (EOM 2). Then, the polarization is analyzed by a polarizing beam splitter (PBS) and a polarization filter in the reflected mode of the PBS, which has a slightly reduced contrast compared to the transmitted mode. After passing through a lens for focusing, the photons are detected on one of two APDs}
\label{fig:QKDsetup}
\end{figure}

The setup is shown in figure \ref{fig:QKDsetup}. The emitted freespace photons from the SPS are initially prepared in a linear polarization state by passing through a linear polarization filter after a $\lambda/2$ plate which is adjusted to maximize the count rate. After passing through a pinhole for further spatial mode cleaning, the photons impinge on the first EOM which is controlled by Alice. The EOM acts as a $\lambda/2$ and $\lambda/4$ plate, respectively, depending on the applied voltage. In this way two orthogonal linearly polarized photon states as well as two orthogonal circular polarization states can be encoded on the incoming photons, compliant to the BB84 protocol \cite{Bennett1984}. After passing through a lens system for recollimation, the photons pass through a second EOM, which is controlled by Bob. Bob randomly chooses a measurement basis by setting the voltage such that the EOM either does not modify the photons or acts as a $\lambda/4$ plate. A circular polarization is thus transformed into a linear one or vice versa. The linear polarization state can then be deterministically analyzed in a system consisting of a polarizing beam splitter (PBS), a linear polarization filter, compensating the non-perfect contrast of the PBS in reflection, and two avalanche photodiodes (APDs) (Perkin Elmer AQR). For the random bit and basis choice of Alice and Bob, quantum random numbers from the online random number service of HU Berlin and PicoQuant GmbH \bibnote [QRNGonline] {http://
qrng.physik.hu-berlin.de/, see also \cite{Wahl2011}} are used.

\section{Experimental results}
In order to test the suitability of two different single photon sources we ran the BB84 protocol in the QKD setup. 

The experimentally obtained parameters are summarized in table \ref{tab:1}. With the brightest NV centre, emitting at a count rate of 8.9\,$kcps$, a sifted keyrate of 4\,$kbit/s$ at a quantum bit error rate (QBER) of 3\,\% was achieved. The raw key was then further processed using the CASCADE protocol \cite{Brassard1994}, resulting in a secure keyrate of 2.6\,$kbit/s$. The CASCADE protocol is an efficient postprocessing protocol for the sifted key aiming at minimizing discrepancies between Alice and Bobs keys while at the same time minimizing Eves possible information about it. It consists of two steps: error correction by parity comparison between Alice and Bob and subsequent dilution of Eves information in the so-called privacy amplification. This privacy amplification is realized by a randomly chosen function belonging to a special class of hash functions. The algorithm which was used here was implemented by ourselves and can be downloaded on our website \bibnote [CASCADEonline] {http://www.physik.hu-berlin.de/nano}.
The observed QBER can be explained by the limited contrast of the polarization optics and the EOMs.\\
Using the brightest stable SiV, emitting at a count rate of 3.7\,kcps, we found a sifted keyrate of 1.5\,$kbit/s$, a QBER of 3\,\% and a resulting secure keyrate of 1\,$kbit/s$. Both keys were transmitted at an experimental clockrate of 1\,MHz.\\ Both NV and SiV centres were running with a stable emission rate over several hours, which was the necessary time to find the right EOM settings for transmission. This shows the long-term stability of our setup.

\begin{table}[h]
\centering
\begin{tabular}{l | l | l }

 & \textbf{NV} & \textbf{SiV} \\ \hline
\textbf{Repetition rate} & 1\,MHz &  1\,MHz \\ \hline
\textbf{Count rate} & 8.9\,kbit/s &  3.7\,kbit/s \\ \hline
\textbf{Sifted keyrate} & 4\,kbit/s & 1.5\,kbit/s \\ \hline
\textbf{QBER} & 3\,\% & 3\,\% \\ \hline
\textbf{Secured keyrate} & 2.6\,kbit/s & 1\,kbit/s  
\label{table1}
\end{tabular}
\caption{Results of the QKD experiments with NVs and SiVs}
\label{tab:1}
\end{table}

\begin{figure}[h] \center
\includegraphics [width=16cm] {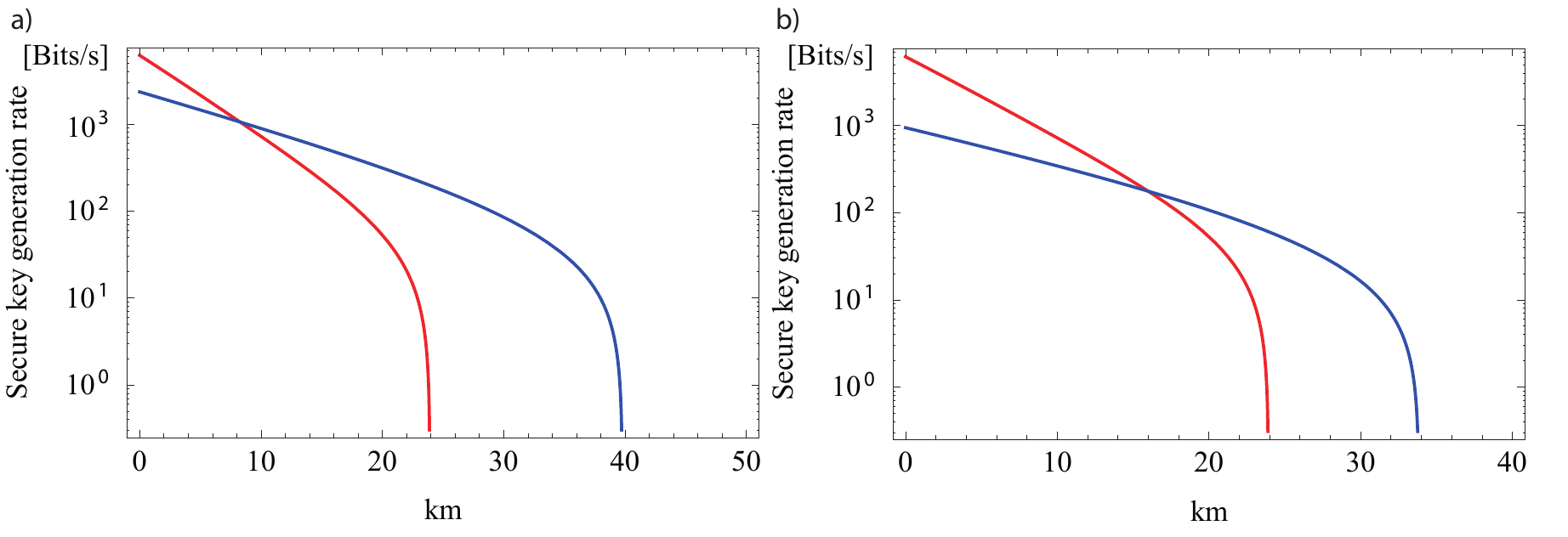}\\
\caption{Secure key rates achieved with the BB84 protocol utilizing a true SPS with an NV centre (a) and a SiV centre (b) in blue. For comparision, the keyrate when using an attenuated laser at the optimized intensity $\mu \sim \eta_{total}$, where $\eta_{total}$ is the overall transmission of the setup, is shown in red.}
\label{fig:rates}
\end{figure}

It is interesting to have a look at the source efficiencies and the $g^{(2)}(0)$ values of the SPS and their consequences on security concerning multiphoton events. An upper bound on the probability $p_{m}$ to have multiphoton events in pulses emitted from a sub-Poisson light sources is given by \cite{Waks2002},

\begin{eqnarray}
p_{m} \leq \frac{\mu^{2} g^{(2)}(0)}{2} ,
\end{eqnarray}

where $\mu$ is the mean photon number per pulse which is identical to the efficiency of the SPS to emit a single photon into the desired mode. This upper bound can then be used to calculate a lower bound on the secure keyrate $R$, which takes the unsecurity of having multiphoton events into account \cite{Gottesman2004}:

\begin{eqnarray}
R\geq q\{-Q_{\mu} f(E_{\mu})H_{2}(E_{\mu}) + Q_{\mu}(1 - \Delta)(1 - H_{2}(E_{\mu}/(1 - \Delta))\}
\label{equ:securerate}
\end{eqnarray}

q is an efficiency factor depending on the exact protocol and expressing the randomness of the base choice, here it is $1/2$, $Q_{\mu}$ is the signal gain of a signal with intensity $\mu$ \bibnote [intensity] {$Q_{\mu}$ is the fraction of detection events at Bob's side which is signal related. For an exact definition, see i.e. \cite{Lo2005}}, $E_{\mu}$ is the error rate of a signal with intensity $\mu$, and $H_{2}(x)$ \bibnote [Entropy] {$H_{2}(x)=-x \cdot log_{2}(x) - (1-x) \cdot log_{2}(1-x)$} is the binary Shannon information function \cite{Shannon1948}. $\Delta$ is the ratio of "tagged" photons. "Tagged" photons are photon qubits emitted by a faulty source, wearing a "tag" with readable information for the eavesdropper revealing the actual basis of the qubit, cf. \cite{Gottesman2004}. When more than one photon is emitted at a time, it carries the same information as its partner photon and can be regarded as a "tagged" photon when exploited by Eve in a PNS attack. For a single photon source $\Delta=p_{m}/p_{click}$, with the detection probability of a signal $p_{click}$ $=$ $\mu \cdot \eta_{total} + p_{dc}$ \cite{Waks2002}. $\eta_{total}$ contains, besides the setup transmission $\eta_{setup}$ (cf. section \ref{subsec:NV}), the distance and medium dependent transmission of the quantum channel. The darkcount probability $p_{dc}$ is in our setup $2.4\times10^{-5}$.
Fig. \ref{fig:rates} shows the results of these secure keyrate calculations as a function of the transmission distance, using a typical transmission of light in air at sea level of 0.4dB/km \cite{Giggenbach2005}. 

The secure key generation rate at 0\,km roughly reproduces our measured keyrate after postprocessing. For comparison we also plotted the secure keyrate of a potential experiment with identical parameters, but with an attenuated laser instead of a single photon source with an intensity of $\mu\sim\eta_{total}$, which is approximately optimal for an attenuated laser without the decoy state protocol and when considering possible PNS attacks \cite{Luetkenhaus2000}. The keyrate is again calculated as in equation \ref{equ:securerate}), taking the multiphoton probability of a Poissonian lightsource into account. At short distances, attenuated laser pulses have relatively high signal intensities and thus outperform our SPS, but at a certain distance (\textgreater 8\,km for the NV and \textgreater 16\,km for the SiV, cf. figure \ref{fig:rates}) the SPS provides higher secure keyrates and longer achievable transmission distances. This is due to a lower number of multiphoton events in a true SPS compared to an attenuated classical pulse of the same intensity. However, for the parameters achieved with single photon sources based on defect centres an attenuated laser together with the decoy state protocol is still favourable over our SPS regarding maximum keyrates and achievable distances. This is because the obtained efficiencies of the SPS is not high enough and at the same time the $g^{(2)}(0)$ value not low enough to outperform WCP together with the decoy state protocol (cf. section \ref{sec:discussion}).

In this context, it is also interesting to note that even SPS sometimes have to be attenuated to reduce the small, but in a real source still existing fraction of multi-photon events to achieve the maximal possible transmission distance \cite{Waks2002}. Even with high source efficiency, security can be compromised at certain distances if the $g^{(2)}(0)$ value is not low enough. In fact for every $g^{(2)}(0)$ value and a given darkcount rate of the detectors, the maximal efficiency to realize the longest distances can be estimated after the following formula, taken from \cite{Waks2002}:

\begin{eqnarray}
\mu_{c} = \sqrt{\frac{2 p_{dc}}{g^{(2)}(0)}}
\label{x}
\end{eqnarray}

The ideal case would be to have a SPS which is above this critical efficiency and to attenuate it in order to achieve the best rates at any distance. For our NV centre, we have an efficiency of 0.029 and a $g^{(2)}(0)$ of 0.09 (cf. section \ref{subsec:NV}). The critical efficiency for this value of $g^{(2)}(0)$ is 0.039, a little above the given efficiency. There are examples of NV centres with higher efficiencies \cite{Schroder2011} which would surpass their critical efficiency. For the SiV and its $g^{(2)}$ value of 0.04, the efficiency of 0.012 (cf. section \ref{subsec:SiV}) is below its critical efficiency of 0.035 due to the low quantum efficiency of the SiV.

\section{Discussion and Conclusion}
\label{sec:discussion}
Our measurements have shown that there are severe constraints regarding the applicability of true single photon sources in QKD applications. Major requirements are a short lifetime, a quantum efficiency near unity, and a strong suppression of multi-photon events. Our QKD testbed allows for long-term stability tests of various room-temperature single photon sources. Due to its versatile design, other defect centres in diamond or more sophisticated photon extraction architectures can be implemented. Examples for the latter are plasmonically enhanced single photon sources \cite{Schietinger2009}, layered dielectric structures \cite{Goetzinger2011} or three-dimensional light guiding microstructures \cite{Schell2013}. Since defect centres in nanodiamonds are point-like sources which can be easily integrated in such structures, collection efficiencies into a usable output mode exceeding 95\,\% are realistic. In this case the internal quantum efficiency would still be a limitation. Recent studies \cite{Frimmer2013,Schell2013_2} have shown that the quantum efficiency may vary significantly within different nanodiamonds, but possibly emitters with more than 90\,\% can be selected. An approach to enhance this requires enhancement of the radiative rate due to the Purcell effect \cite{Wolters2010,Riedrich2012} which is technically more challenging for a room-temperature emitter \cite{Albrecht2013}. A suppression of multi-photon events should be easier if the emission is narrow band (e.g. as in the SiV or Cr centre compared with the NV centre) and allows for better filtering of the signal against the background. 

\begin{figure}[h] \center
\includegraphics [width=12cm] {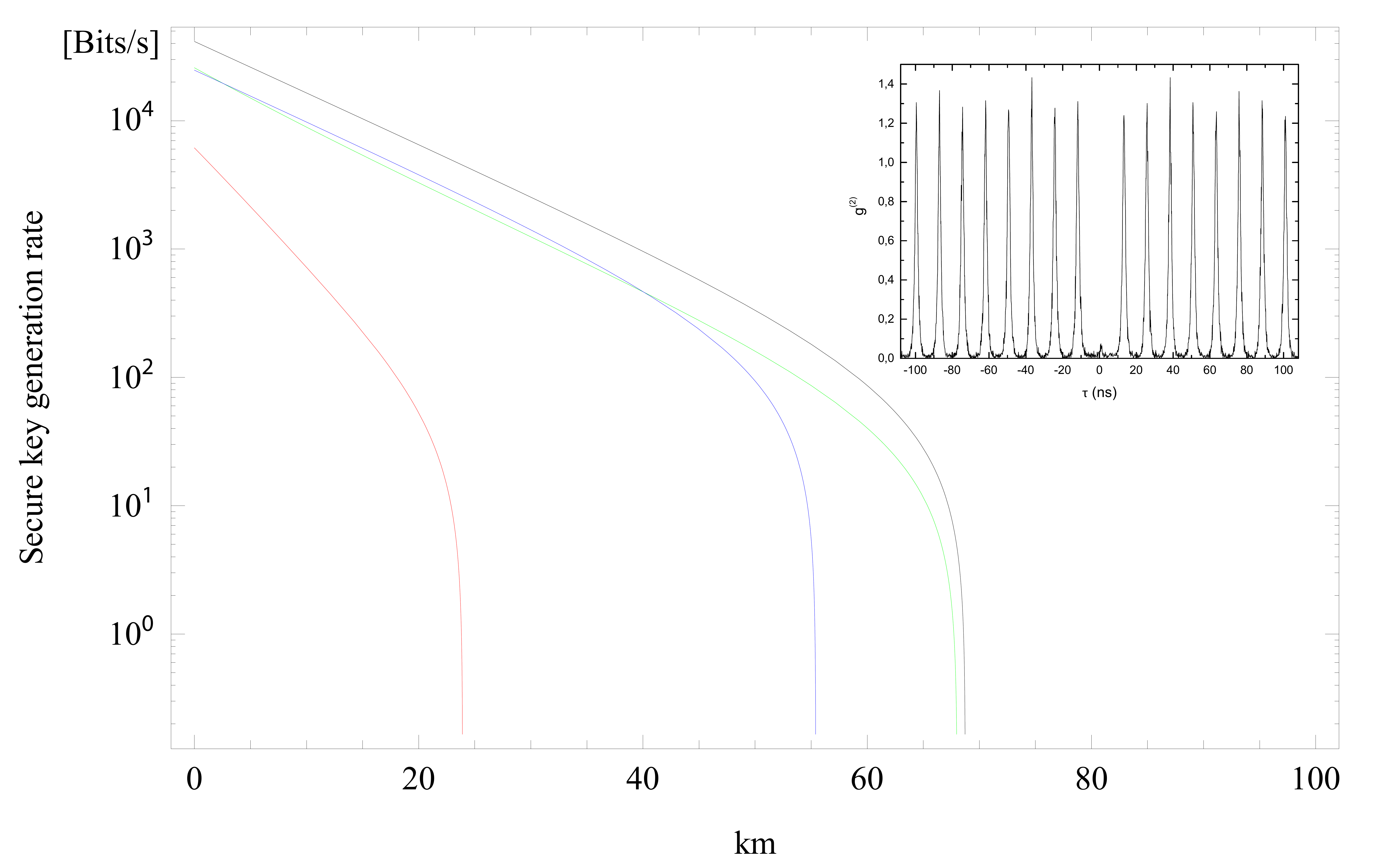}\\
\caption{Here, the secure key rates of defect centres with ideal, realistic features are depicted. For these sources, quantum efficiencies of about unity and photon yields of 10\,\% (blue curve) and 95\,\% (black curve) are assumed. The $g^{2}(0)$ value is 0.005 (blue curve) and 0.0005 (black curve). For comparison, an attenuated laser without (red curve) and with (green curve) decoy state protocol is depicted as well. The inset shows the $g^{(2)}$ value for a SiV for pulsed excitation at a repetition rate of 80\,MHz. The countrate is 230\,kcps with $g^{(2)}(0)$ $<$ 0.1}
\label{fig:rates_ideal}
\end{figure}

In figure \ref{fig:rates_ideal} we estimated the maximum possible secure key generation rates for a next generation single photon source based on defect centres.

As already mentioned, the SiV would allow for much higher excitation rates than shown here, which would lead to much higher keyrates. The inset of figure \ref{fig:rates_ideal} shows the $g^{(2)}$ value for an SiV for pulsed excitation at a repetition rate of 80\,MHz, featuring a $g^{(2)}(0)$ $<$ 0.1 at a countrate of 230\,kcps.    

Figure \ref{fig:rates_ideal} demonstrates that single photon sources based on defect centres in diamond could indeed outperform attenuated light sources. Since all these sources have excellent optical properties and operate at room temperature it can be expected that they will also be accepted in commercial devices. 

In summary, we implemented a QKD experiment with a compact SPS setup using defect centres. The setup of the source is such that it can easily be deployed in various QIP experiments. NV and SiV centres are used and can easily be interchanged within the experiment. SiV centres are potentially interesting for QIP applications due to their narrow spectral emission bandwidth but still suffer from drawbacks concerning their efficiency and their stability, which have to be overcome in the future. A thorough security analysis of the QKD experiment, taking the source efficiency and its $g^{(2)}(0)$ value into account, is established.

\ack
Funding by BMBF (KEPHOSI) and DFG (SFB787) is acknowledged by HU Berlin. Christoph Becher and his co-workers from Universit\"{a}t des Saarlandes thank the Bundesministerium f\"{u}r
Bildung und Forschung within the projects EphQuaM (contract 01BL0903) and QuOReP (contract 01BQ1011) for their financial support. We thank M. Fischer, S. Gsell and M. Schreck (University of Augsburg) for supplying the CVD nanodiamond samples containing SiV centres. T. Schr\"{o}der also acknowledges support by the Alexander von Humboldt Stiftung.

\section*{References}

\bibliography{Diamonds_QKD_NJP}{}
\bibliographystyle{unsrt}


\end{document}